\documentclass[onecolumn,aps,prr,preprintnumbers,amsmath,amssymb]{revtex4-2}

\usepackage{natbib}
\usepackage{graphicx}
\usepackage{bm}
\usepackage{color}

\usepackage{amsmath}
\usepackage{amsfonts}
\usepackage{amssymb}
\usepackage{makeidx}
\usepackage{graphicx}

\begin{document}
	
	\title{Dissipative dynamics of an interacting spin system with collective damping}
	\author{Irfan A Dar$^1$, Faisal Farooq$^1$, Junaid Majeed$^2$,
Mehboob Rashid$^3$, Sheikh Irfan$^1$, Muzaffar Qadir Lone$^1$\footnote{corresponding author: lone.muzaffar@uok.edu.in}}
	\affiliation{$^1$Quantum Dynamics Lab, Department of Physics, University of Kashmir, Srinagar-190006 India.\\
	 $^2$International Center for Theoretical Physics,Bengaluru - 560 089, India.\\
	$^3$National Institute of Technology,  Srinagar-190006, India}  
	
\begin{abstract}
	The competition between Hamiltonian and Lindblad dynamics in quantum systems give rise to non-equillibrium phenomena with no counter part in conventional condensed matter physics. In this paper, we investigate this interplay of dynamics in infinite range  Heisenberg model coupled to a non-Markovian bath and subjected to Lindblad dynamics due to spin flipping at a given site. The spin model is bosonized via  Holstein-Primakoff transformations and is shown to be valid for narrow range of parameters in the thermodynamic limit. Using Schwinger-Keldysh technique, we derive mean field solution of the model and observe that the system breaks    $\mathcal{Z}_2$-symmetry at the transition point. We calculate effective temperature that has linear  dependence
	on the effective system-bath coupling, and is
	independent of the dissipation rate and  cutoff  frequency  of the bath  spectral density.  Furthermore, we study the fluctuations 
	over mean field  and  show that  the dissipative spectrum  is modified by  ${\rm O}(\frac{1}{N})$ correction term
	which results change in various physically measurable quantities.
	%
\end{abstract}

\maketitle

\section{Introduction}

The success of equilibrium statistical mechanics in describing thermal properties of matter can be gauged through two of its robust predictions- the emergent phenomena and universality\cite{1,2}. However,
recent experiments  ranging from polariton condensates in context of semiconductor quantum wells in optical cavities\cite{3,4,5}, arrays of microcavaties\cite{6} to trapped ions\cite{7,8}, optomechanical setups\cite{9,10} and strongly
interacting Rydberg polaritons\cite{11,12} to explore the bulk behaviour of ultracold matter in presence of drives and
dissipiations lead to the breakdown of traditional equilibrium techniques. At microscopic scale, the very
symmetry responsibe for implementing the thermal order is broken down due to presence of drives and
this breakdown is manifested in resulting breakdown of detailed balance principle\cite{13,14,15}. In addition to coherant
dynamics goverened by Hamiltonian the above mentioned systems are also driven by dissipation requiring
non-conventional evolution schemes characterised by competition between drives and dissipiations. The
many-body stationary states of novel evolutionary schemes emerge as new non-equilibrium phases of
matter\cite{16,17,18,19,20}. At macroscopic scale,in order to fix the notion of universality for emergent non-
equilibrium phases, the fundamental challenge remains to look for alternatives for equilibrium notions
like temperature, free energy and entropy which become vaguely defined for non-equilibrium phases of
matter\cite{21}. So the analogs of equilibrium notions like temperature etc.
need to emerge self-consisently as a result of dynamics of the model\cite{20,22,23,24,25,26,27,28,29,30}. Therefore, despite non-equilibrium 
ingredients these systems equilibrate effectively and as a result of competition between drives and
dissipations effective temperature is identified with suitable combination of dissipative parameters of
underlying model via fluctuation-dissipation relations. However, despite effective equilibration these
systems also defy traditional equilibrium signatures measured through corresponding response functions
The driven open quantum systems can be well described by microscopic master equations\cite{31,32}, but the
traditional techniques of quantum optics cannot be used efficiently We employ Lindblad Master
equation and map it to the Schwinger-Keldysh (SK) functional integral  formalism\cite{14,18,33} to study the
dissipative dynamics in an interacting spin model with long range interactions described by an anisotropic
Heisenberg model\cite{34,35} and coupled to a non-Markovian bath. This approach has found numerous applications
to driven-dissipative systems such as lossy polariton condensates\cite{18,19,36} and driven atomic ensembles
interacting with a cavity mode\cite{20}.

In general, a system of qubits can be represented by some interacting spin-half particles. The interaction between these qubits can be nearest neighbor on a given lattice\cite{35} or fully connected in a sense all spins interact with each other\cite{37}. In this work, we consider a fully connected model described by anisotropic antiferromagnetic Heisenberg model (IRHM) where each spin is coupled to a same  bosonic bath. We also consider Lindblad dynamics via a collective  spin flipping. Lindblad dynamics is essentially Markovian in nature and our aim is  to understand the influence of both Markovian and non-Markovian effects on the underlying dynamics of the system\cite{22}.

The long range interactions are used to study the fully connected in the context of  light harvesting complexes\cite{38,39}. An example of such model is Fenna-Mathews-Oslo model\cite{38,39,40} for excitons, where the hopping energy is uniform and the system bath coupling is not very weak but of intermediate range\cite{37,40,41}. Furthermore,  these interactions can be produced in cavity quantum electrodynamics experiments\cite{31,42,43}.

This paper is organized as follows. In section II, we introduce the IRHM coupled with a bosonic bath. 
Using Holstein-Primakoff (HP) transformations, we bosonize IRHM model and  map it to a self interacting bosonic mode and find the parameter values where HP transformation breaks down.
The complete Hamiltonian becomes Dicke model with non-linearities. In next section III, we make use 
of SK functional integral formalism to study the steady state solutions of the equations of motion.
We see that the critical coupling depends on the spectral density of the bath. 
In section IV, we study the dissipative spectrum beyond mean field level and analyze the effect of  fluctuations 
on different observables. Finally,  we conclude
in section V.
\section{Bosonization of IRHM coupled with bosonic bath}
We consider a fully connected model of qubits represented by 
spin-$\frac{1}{2}$ particles interacting with each other  through 
a infinite range   Heisenberg antiferromagnetic exchange interaction $H_{\rm IRHM}$\cite{34,37,41} with anisotropy in the longitudinal channel, and 
coupled to a non-Markovian bath:
\begin{eqnarray}
	H= H_{\rm IRHM}
	+ \sum_k \omega_k b_k^{\dagger} b_k +  \frac{1}{\sqrt{N}} \sum_{i,k}S_i^x(g_k b_k + g_k^* b_k^{\dagger})
	\label{th}
\end{eqnarray}
where
\begin{eqnarray}
	H_{\rm IRHM} = \frac{J }{N}\sum_{i,j>i}  \left [  S^x_i S^x_j + S_i^y S_j^y +\Delta S^z_iS^z_j \right]
	\label{Hs}
\end{eqnarray} 
where $J>0$ sets the energy scale of the model, and $\Delta$ represents  an anisotropic parameter. $\vec{S}=\frac{\hbar}{2}\vec{\sigma}$, $\vec{\sigma}$ are Pauli matrices and $S_{\pm} = S^x \pm i S^y$ are ladder operators. We note that $[H_{\rm IRHM},\sum_i S^z_i ]=0$ and $[H_{\rm IRHM}, S^2_{Total} ]=0$, so that the eigen states of $H_{\rm IRHM}$ are described by total $S_T$ and $S_Z$ values. The ground state corresponds to singlet state $S^z_T =0$ and $S_T=0$, which has been shown to form a new class of highly entangled resonating valence bond states. This model has its own importance in describing zigzag graphene nanodisc\cite{35,44,45,46,47,48}, Lipkin-Meshkov-Glick model \cite{49}for certian parameter range. 
For quantum computation and information using quantum dots, the spin states are being prepared,manipulated, and measured using rapid control of Heisenberg exchange interaction\cite{50,51,52}.

Next we define total spin operator $\Vec{S}=  \sum_i \Vec{S}_i$  and
bosonize the $H_{\rm IRHM}$  using
Holstein-Primakoff transformations \cite{53}:
\begin{eqnarray}
	S^{+}&=&  \sqrt{N-a^{\dagger} a}~ a\\
	S^{-}&=&  a^{\dagger}\sqrt{N-a^{\dagger} a}\\
	S^z &=&   \frac{N}{2}-a^{\dagger} a
\end{eqnarray}
with $N$ as the total number of spin-1/2 particles.
Therefore, $H_{\rm IRHM}$ is mapped to a bosonic mode with non-linearities at various orders  of $\frac{1}{N}$:

\begin{eqnarray}
	H_a = \frac{J}{2}(1-\Delta-\frac{1}{N})a^{\dagger}a + \frac{J}{2N}(\Delta-1 +\frac{1}{2N}) (a^{\dagger}a)^2 + \frac{J}{4N^2} (a^{\dagger}a)^3+....
	\label{Ha}
\end{eqnarray}
In the thermodynamic limit, we see that the $H_a \sim  \frac{J}{2}(1-\Delta) a^{\dagger}a$, 
$\Delta=1$ therefore, breaks the validity of the HP transformation as $H_a$ vanishes in this limit, and at this point ground state becomes degenerate at lowest order of perturbation. Restricting to the case of $\Delta<1$ and retaining the $O(\frac{1}{N})$ terms,  we see that the coefficient of quartic term $(a^{\dagger}a)^2$ becomes negative for large $N$ implying the instability of the bosonic mode. Therefore, for the stable ground state in the finite $N$ limit, we require sextic term which is of the $O(\frac{1}{N^2})$ with positive coefficient (eqn. \ref{Ha}).  Next, we see that for  $J\Delta>>1$, even the coefficient of quartic term becomes positive, the system does not have a stable ground state in thermodynamic limit. This breakdown of HP transformation can mainly be attributed to different types of phase transitions occurring in the model as we vary $\Delta$ from $-\infty$ through 0 to $+\infty$; and to distance independent nature of the interactions in contrast to nearest neighbor $XXZ$
model. Furthermore, we are interested in effect of bath in the thermodynamic limit of the model and it suffices to take $0<\Delta<1$. 
Therefore, we write 
the total Hamiltonian given by  equation \ref{th} as 
\begin{eqnarray}
	H_{eff} = H_a
	+ \sum_k \omega_k b_k^{\dagger} b_k +   \frac{1}{2}\sum_{k}(a+a^{\dagger})(g_k b_k + g_k^* b_k^{\dagger})
	\label{bosonize}
\end{eqnarray}
This is just Dicke model \cite{54} with non-linearities, and couped to a multimode bath.
The above model possess $\mathcal{Z}_2$-symmetry. In the strong coupling regime within thermodynamic limit,  the ground state
of the above model breaks this $\mathcal{Z}_2$-symmetry and exhibits a phase transition to phase with $\langle a\rangle\neq0 $.

Next, we assume a dissipative process in addition to the 
coherent dynamics represented by the Hamiltonian in equation \ref{bosonize}, 
due to spin flipping (spontaneous emission) at site $i$ 
from $|\uparrow \rangle $ to $|\downarrow\rangle$ at a rate of $\kappa$,  represented by 
Lindblad master equation:
\begin{eqnarray}
	\frac{d\rho_s}{dt} &=& -i[H_{\rm IRHM}, \rho_s]+ \frac{\kappa}{N} \sum_{i,j}[2 S^+_i \rho_s S_j^- -\{S^+_i S^-_j,\rho_s\}] \\
	&=& -i[H_a, \rho_s]
	+ \kappa [2 a \rho_s a^{\dagger}  -\{a^{\dagger} a,\rho_s\}]
\end{eqnarray}
where in second line we have used Holstein-Primakoff transformations, $\rho_s$ is the density matrix corresponding to 
$a$-fields.

\section{Schwinger-Keldysh Field Theory}
In this section we use SK field theoretic technique to study the dynamics in the model considered. The SK field theory is the path integral representation of the time evolved density matrix $Z={\rm Tr} \rho(t)$  on a closed time contour with fields defined along two branches called as forward (backwrd )time branch $(\pm)$, such that both branches meet at $t=\infty$. The partition function $Z$ can be therefore written for some field $\phi(x)$ as
(with $\bar{\phi}$ as the conjugate of $\phi$,  )as

\begin{eqnarray}
	Z=\int D[\bar{\phi}_+,\phi_+,\bar{\phi}_-,\phi_-] e^{iS_{SK}[\bar{\phi}_+,\phi_+, \bar{\phi}_-, \phi_-]},
\end{eqnarray}
where Schwinger-Keldysh action for the total system plus bth including Lindblad dynamics is $S_{SK}=S_0 + S_D$. $S_0$ is action for Hamiltonian dynamics
\begin{eqnarray}
	S_{0}= \sum_{\eta =\pm} \eta \int dx dt ~[\bar{\phi}_{\eta} i\partial_t \phi_{\eta} -H(\bar{\phi}_{\eta},\phi_{\eta})].
\end{eqnarray}
The action  corresponding to dissipation due to Lindblad operator is given by $S_D$
\begin{eqnarray}
	S_D= -i\kappa\int dx dt ~2\phi_+ \bar{\phi}_- -(\bar{\phi}_+ \phi_+ +\bar{\phi}_- \phi_- ).
\end{eqnarray}

Therefore, for the model under consideration, we  write SK action as
$S_{SK}=S_a + S_b+ S_{ab}$. For  $a$-type fields $(S_a$ as action) we have
\begin{eqnarray}
	S_a= \int \!dt \Bigg[ \sum_{\sigma=\pm} \sigma  [\bar{\phi}_{\sigma}(i\partial_t-\omega_0)\phi_{\sigma} 
	+ \frac{\lambda}{N} (\bar{\phi}_{\sigma} \phi_{\sigma})^2 +\frac{J}{4N^2}(\bar{\phi}_{\sigma} \phi_{\sigma})^3 ] -i\kappa ( 2 \phi_+ \bar{\phi}_- - \bar{\phi}_+ \phi_+ - \bar{\phi}_- \phi_-  ) \Bigg].
\end{eqnarray}
Here  $\omega_0= \frac{J}{2}(1-\Delta-\frac{1}{N})$ and $\lambda=\frac{J}{2}(-1+\Delta + \frac{1}{2N})$. $\phi$ represents the bosonic coherent state of $a$-type bosons, $\bar{\phi}$ represents the complex conjugate of $\phi$.
Plus (minus)  signs refers to the field defined on forward (backward) branch of Keldysh contour.
Similarly, if $\psi$ represents the bosonic coherent state of $b$-type bosons ($S_b$ as action), we can 
write 
\begin{eqnarray}
	S_b &=& \int \!dt \sum_k  \sum_{\sigma=\pm} \sigma  [\bar{\psi}_{k\sigma}(i\partial_t-\omega_0)\psi_{k\sigma} \nonumber \\
	S_{ab} &=& -\frac{1}{2}  \int \!\!dt\sum_k g_k \!\! \sum_{\sigma=\pm} \sigma (\bar{\phi}_{\sigma} + \phi_{\sigma}) ( \bar{\psi}_{k\sigma} + \psi_{k\sigma}  )
\end{eqnarray}

Next we implement Keldysh rotation defined as:
$
	\phi_{cl} = \frac{\phi_+ + \phi_- }{\sqrt{2}},
	\phi_q = \frac{\phi_+ - \phi_- }{\sqrt{2}}
$
The subscripts $cl$
and $q$ stand for the classical and the quantum components of the fields, respectively, because the first one can acquire
expectation value while the second one cannot. In this basis, with 
the same transformations for $\psi_k$-field  as well,
we get

\begin{eqnarray}
	S_a &=& \int \!\!dt \Bigg[\begin{pmatrix} \bar{\phi}_{cl}(t)& \bar{\phi}_q(t) \end{pmatrix}
	\begin{pmatrix}
		0& i\partial_t - \omega_0 -i\kappa\\
		i\partial_t - \omega_0 +i\kappa & 2i\kappa
	\end{pmatrix}
	\begin{pmatrix}
		{\phi}_{cl}(t)\\
		\phi_q(t)
	\end{pmatrix} \nonumber\\ 
	&&~~~~~~~+ \frac{\lambda}{2N}  (|\phi_{cl}|^2 + |\phi_q|^2)(\bar{\phi}_{cl} \phi_q + \phi_{cl} \bar{\phi}_q)+ 
	\frac{J}{4\sqrt{2}N^2}\left[(\bar{\phi}_{cl} \phi_q)^3 
	+(\bar{\phi}_{q} \phi_{cl})^3\right.\nonumber \\
	&&~~~~~~~\left. + 3( |\phi_{cl}|^4+ |\phi_q|^4+3|\phi_{cl}|^2|\phi_q|^2)(\bar{\phi}_{cl} \phi_q + \phi_{cl} \bar{\phi}_q )\right]\Bigg]\\
	S_b &=& \sum_k\int \!\!dt \begin{pmatrix} \bar{\psi}_{k cl}(t)~~  \bar{\psi}_{kq}(t) \end{pmatrix}
	\begin{pmatrix}
		0& i\partial_t - \omega_k -i\epsilon\\
		i\partial_t - \omega_k +i\epsilon & 2i\epsilon
	\end{pmatrix}
	\begin{pmatrix}
		{\psi}_{kcl}(t)\\
		\psi_{kq}(t)
	\end{pmatrix}\\
	S_{ab} &=& -\frac{1}{2} \sum_k g_k\int \!\!dt \Bigg[  (\bar{\phi}_{cl} + \phi_{cl}) ( \bar{\psi}_{k q} + \psi_{k q}  )   
	+ ( \bar{\psi}_{k cl} + \psi_{k cl}  )(\bar{\phi}_{q} + \phi_{q}) \Bigg]
\end{eqnarray}
where $\epsilon$ is the regularization parameter. Markovian dissipation is idenfied by the frequency independent part of Keldysh component\cite{20}.
Next we perform saddle point approximation by varying action $S$ with respect to quantum component
of the fields,i.e. $\frac{\delta S}{\delta \bar{\phi}_q}=0$ and $\frac{\delta S}{\bar{\psi}_{kq}}=0$ at
$\phi_{cl}=\phi_0,~ \phi_q=0$ and $\psi_{kcl}=\psi_{k0},~ \psi_{kq}=0$
and get
\begin{eqnarray}
	(-\omega_0+ i\kappa) \phi_0 + \frac{\lambda}{2N}|\phi_0|^2 \phi_0  +\frac{3J}{4\sqrt{2}N^2}|\phi_0|^4\phi_0- \frac{1}{2}\sum_k g_k (\bar{\psi}_{k0} + \psi_{k0})&=&0 \nonumber\\
	(-\omega_k + i\epsilon) \psi_{k0} -\frac{1}{2}g_k (\bar{\phi}_0 + \phi_0)&=&0 \nonumber\\
	\label{saddle}
\end{eqnarray}

In order to solve above equations, we  define bath spectral density $J(\omega)=\sum_k g_k^2 \delta(\omega-\omega_k)$. We consider
the following general form of $J(\omega)$ with Drude-Lorentz cutoff:
\begin{eqnarray}
	J(\omega) =2\pi \gamma \omega \Bigg(\frac{\omega}{\Omega}\Bigg)^{s-1} \frac{\Omega}{\omega^2 + \Omega^2}
	\label{ohm}
\end{eqnarray}
with $\gamma$ as the effective coupling between system and bath, $\Omega$ is the cutoff frequency. $s=1$ correspond to Ohmic
bath, $0<s<1$ and $s>1$ are called
sub-ohmic and super-ohmic baths respectively. However, we will work with ohmic bath $s=1$ for simplicity.
Using this form of spectral density, we see that the saddle point equations \ref{saddle} at $O(\frac{1}{N})$ admit  a
trivial solution $\phi_0 =0$ for symmetric state
and a non-trivial
solution $\phi_0 \ne 0$ for symmetry broken state
which is given by
\begin{eqnarray}
	|\phi_0| = \pm \sqrt{\frac{N\pi}{\lambda}} \Bigg(\gamma_0 -\gamma\Bigg)^{\frac{1}{2}}
\end{eqnarray}
where $\gamma_0 = \frac{1}{\pi} \frac{\omega_0^2 + \kappa^2}{\omega_0}$ is the critical coupling.

Now we evaluate the various correlation function corresponding to $\phi$-field within the mean field level. 
In the thermodynamic limit $N\rightarrow \infty$, the contribution from $O(1/N)$ terms can be ignored. We
first eliminate the $\psi$-field using Gaussian integration. 
Defining $ \Phi_{cl/q} = \begin{pmatrix}
	\phi_{cl/q}(\omega) \\
	\bar{\phi}_{cl/q}(-\omega)
\end{pmatrix} $
and $ \Psi_{cl/q} = \begin{pmatrix}
	\psi_{cl/q}(\omega) \\
	\bar{\psi}_{cl/q}(-\omega)
\end{pmatrix} $
such that  Keldysh-Nambu spinor is defined as
$\eta_8(\omega) = [ \Phi_{cl} ~  \Psi_{kcl}~   \Phi_{q}~ \Psi_{kq} ]^T $.
Using the notation $\int_{\omega} = \int_{-\infty}^{\infty} \frac{d\omega}{2\pi}$ and
$\phi_{cl/q}(t) =\int_{\omega} e^{-i\omega t} \phi_{cl/q}(\omega)$ as the Fourier transform of the $\phi$-field,
we integrate out $\psi$-field to get the following effective action for $\phi$-field:

\begin{eqnarray}
	S_{\rm eff} = \int_{\omega} \eta_4^{\dagger}(\omega) \begin{pmatrix}
		0 & [G^A_{2\times 2}]^{-1}(\omega) \\
		
		[G^R_{2\times 2}]^{-1}(\omega)& D_{2\times 2}^K      \end{pmatrix} \eta_4(\omega)
	\label{Sef}
\end{eqnarray}

where $\eta_4(\omega) = \begin{pmatrix}\Phi_{cl}(\omega) \\ \Phi_{q}(\omega) \end{pmatrix}$ ,
$ D^K_{2\times 2} = {\rm diag}(2i\kappa,2i\kappa)$.
The retarted Green's function is given by
\begin{eqnarray}
	[G^R_{2\times 2}]^{-1}(\omega) =\begin{pmatrix}
		\omega-\omega_0 + i\kappa + \Sigma^R(\omega) & \Sigma^R(\omega)\\
		[\Sigma^{R}(-\omega)]^*& -\omega-\omega_0 - i\kappa + [\Sigma^R(-\omega)]^*
	\end{pmatrix}. \nonumber \\
\end{eqnarray}
Here $\Sigma^R(\omega) =[\Sigma^{R}(-\omega)]^* = -\frac{1}{2}\sum_k\frac{|g_k|^2\omega_k}{\omega^2-\omega_k^2} $ is 
the self energy function. Thus it is evident that self energy depends on the density of bath states.Using
the density of states given by equation \ref{ohm}, we write the  self-energy function $\Sigma(\omega)  \equiv\Sigma^R(\omega) $ for 
Ohmic case as
\begin{eqnarray}
	\Sigma(\omega) = \frac{\pi}{2} \gamma \frac{\Omega^2}{\omega^2 + \Omega^2}
\end{eqnarray}

The characteristic frequencies of the system are defined by the zeros of the determinant $[G^R_{2\times 2}]^{-1}(\omega)$ those 
correspond to the poles of the response function $G^R_{2\times2}(\omega)$. Since Green's function possess the symmetry that
$\sigma_x  G^R_{2\times 2}(\omega) \sigma_x =  [G^R_{2\times 2}(-\omega)]^{\star} $, so that 
the roots come into pairs with  opposite real parts or are purely imaginary.
Thus the dispersion of dissipative modes are given by 
${\rm det} [G^R_{2\times 2}]^{-1}(\omega) =0$ which implies
\begin{eqnarray}
	\omega =- i\kappa  \pm \sqrt{\omega_0^2- 2\omega_0 \Sigma(\omega) } 
	\label{char}
\end{eqnarray}

\begin{figure*}[t] 
	\begin{center}
		{\includegraphics[width=1.8in,height=2in]{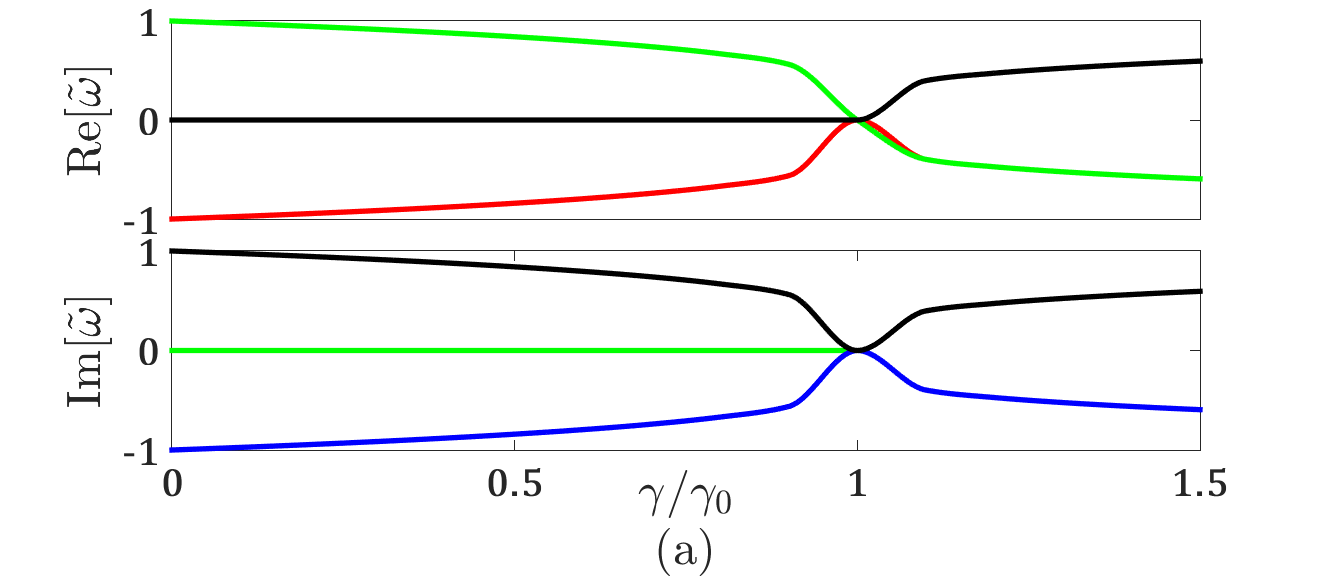}}\hspace*{5pt}
		{\includegraphics[width=1.8in,height=2in]{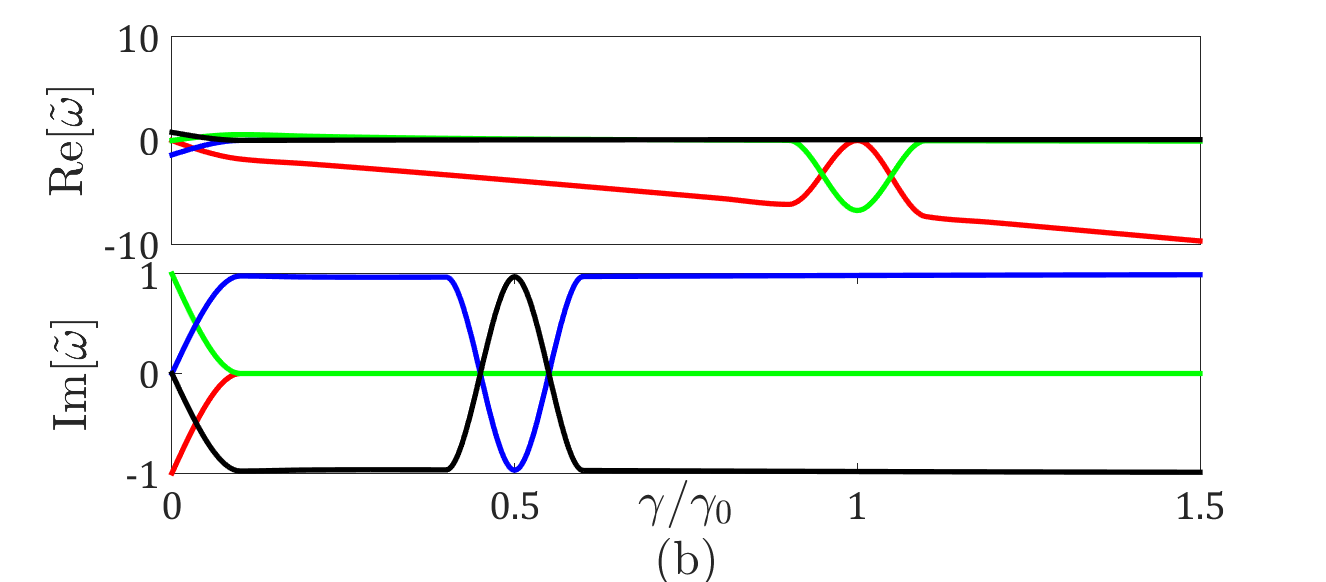}}\hspace*{5pt}
		{\includegraphics[width=1.8in,height=2in]{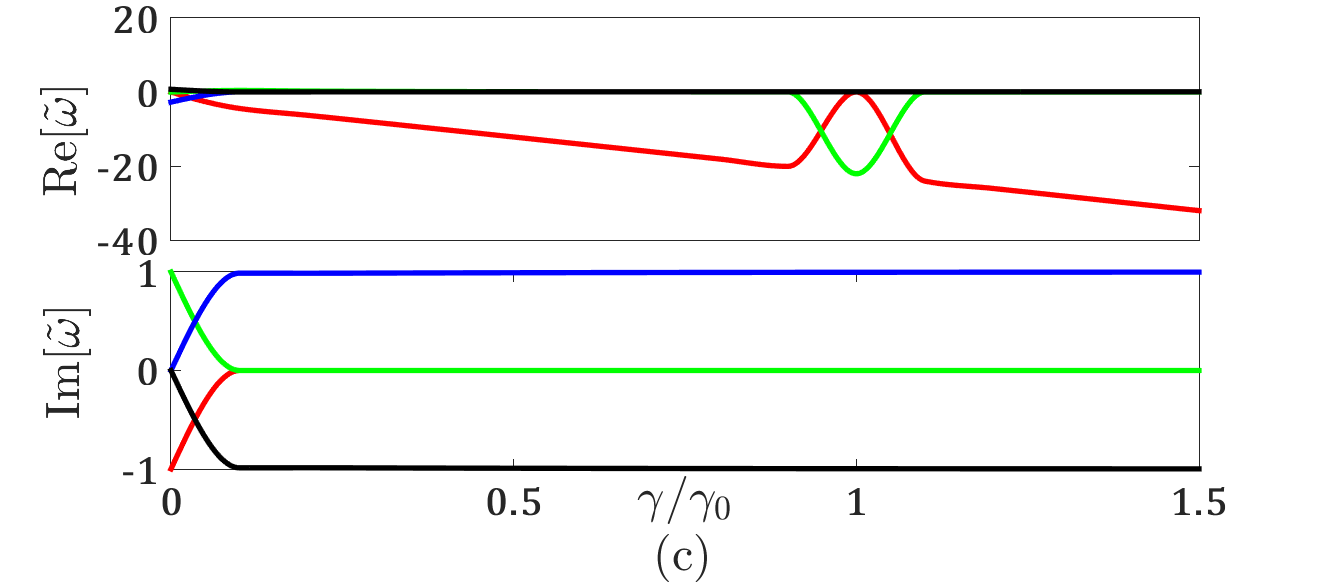}}
	\end{center}
	\caption{Real and imaginary part of the roots of the equation \ref{char} pertaining to characteristic frequencies of the system
		$\tilde{\omega}=\frac{\omega}{\Omega}$.  The values of $k$ chosen are (a) $k=0$, (b) $k=0.3$ and (c) $k=1$. 
		We see that  one of the eigen modes (red curve) vanish at $\gamma=\gamma_0$ for some finite $k$ value.}
	\label{Fig1}
\end{figure*}

Figure \ref{Fig1} is the plot of real and imaginary parts of the roots of the above characteristic equation for 
different values of $k$ with anisotropic parameter $\Delta= 0.7$, $J=1$. We see that for no spin-flipping case $k=0$, we have all the roots vanishing at
transition point $\frac{\gamma}{\gamma_0}=1$ as expected. As we increase value of $k$, 
different modes hybridize and get shifted in the opposite directions . On approach to transition point two solutions
become purely imaginary and correspond to damped modes as shown
by  blue and black curves in 
the figure \ref{Fig1}(b) $\&$ (c) .  While at transition point  only one mode shown by red  
curve in figure \ref{Fig1} (b) $\&$ (c) vanish and thus making the system dynamically unstable.

\subsection{Correlation Functions}

\begin{figure}[t] 
	\includegraphics[width=3in,height=3.0in]{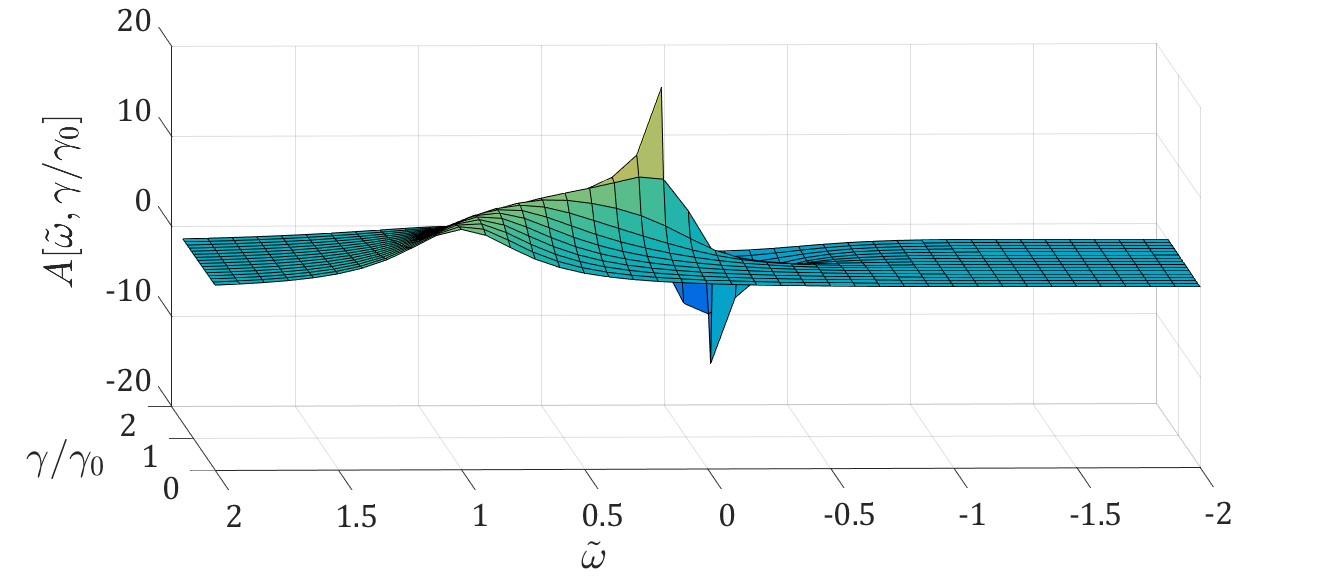} \hspace*{5pts}
	\includegraphics[width=3in,height=3.0in]{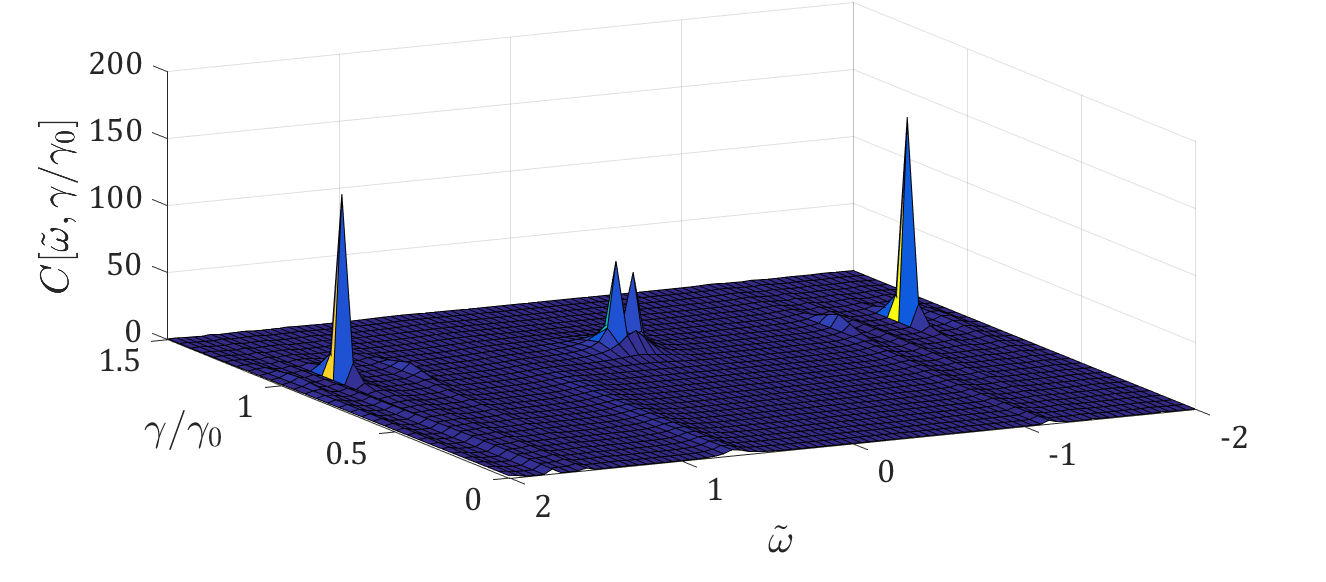}
	\caption{left:Spectral response function $A(\tilde{\omega},\frac{\gamma}{\gamma_0})$ as a function of $\frac{\gamma}{\gamma_0}$ and $\tilde{\omega} = \frac{\omega}{\Omega}$ 
		for $k=0.3$ , $\omega_0=1$. The Lorentzian peak at $\gamma=0$ is shifted towards low frequency mode at transition point.Right: Correlation fucntion $C(\tilde{\omega},\frac{\gamma}{\gamma_0})$
		as a function of $\frac{\gamma}{\gamma_0}$ and $\tilde{\omega} = \frac{\omega}{\Omega}$ 
		for $k=0.3$, $\omega_0=1$.}
	\label{Fig2}
\end{figure}

The phyically measurable quantities are correlation functions. The spectral response function $A(\omega)$ encodes the 
systems response to the active, external perturbations. It is defined as
\begin{eqnarray}
	A(\omega) = i[G^R(\omega)-G^A(\omega)].
\end{eqnarray}
In the present case, we write $A(\omega) = -2{\rm Im} G^R(\omega)$ and is given by
\begin{eqnarray}
	\!\!\!\!\!\!\!
	A(\omega) = \frac{2[ (\omega^2 + \kappa^2+\omega_0^2 + 2 \omega \omega_0 )\kappa - 2\kappa(\omega_0 + \omega)\Sigma]}
	{(\omega^2 -\kappa^2 -\omega_0^2 + 2 \omega_0 \Sigma)^2 + 4 \omega^2 \kappa^2 }
	\label{spec}
\end{eqnarray}
At $\gamma=0$, we see from the figure \ref{Fig2} that $A(\omega)$ has Lorentzian shape centered at $\omega_0$. As $\gamma$ increases towards 
$\gamma_0$, the Lorentzian peak gets shifted towards low frequency mode $\omega=0$ at transition point.

The correlation function encodes the systems internal correlations and is defined as
\begin{eqnarray}
	C(t,t^{\prime}) =\langle \{\hat{a}(t), \hat{a}^{\dagger}(t^{\prime})\} \rangle= i G^K(t,t^{\prime})  
\end{eqnarray}
In steady state, we write
\begin{eqnarray}
	C= 2\langle a^{\dagger} a\rangle + 1 = i\int \frac{d\omega}{2\pi} G^K(\omega)
	\label{corr}
\end{eqnarray}
with 
\begin{eqnarray}
	iG^K (\omega) = \frac{2\kappa[(\omega+ \omega_0 -\Sigma)^2 + \kappa^2 + \Sigma^2 ]}{(\omega^2 -\kappa^2 -\omega_0^2 + 2 \omega_0 \Sigma)^2 + 4\omega^2 \kappa^2}
\end{eqnarray}

\begin{figure}[t] 
	\includegraphics[width=3.0in,height=2.0in]{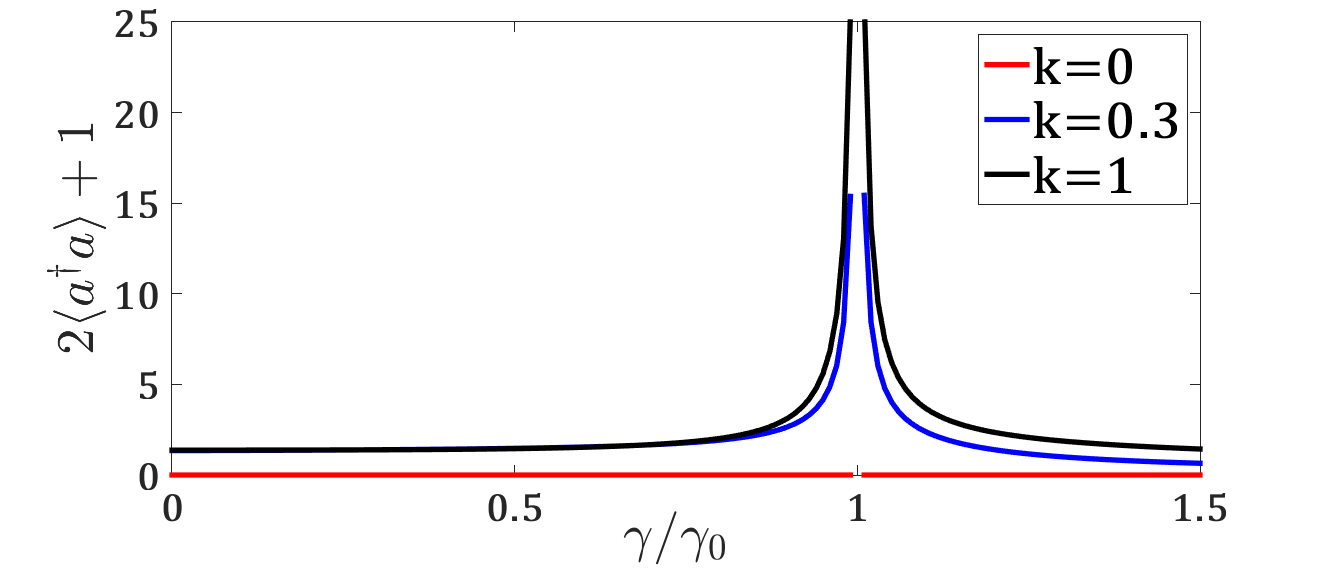}
	\caption{Steady state  number density for different values of $\kappa$ and  $\omega_0=1$. The distribution funtion diverges as $|\gamma-\gamma_0|^{-\alpha}$ with $\alpha=0$ for $\kappa=0$ and $\alpha=1$ for $\kappa\ne0$.  }
	\label{Fig4}
\end{figure}

For a decaying bosonic mode with no coupling to the bath i.e. $\gamma=0$, we see 
from the equations \ref{spec} and \ref{corr} that $C(\omega)=A(\omega)$, and 
the steady state boson density $\langle a^{\dagger} a \rangle=0$, which corresponds to the vacuum of the $\phi$-field. 
We see from the figure \ref{Fig2} that  there occurs divergence $C(\tilde{\omega})$ for $\tilde{\omega}=0$ at transition 
point $\frac{\gamma}{\gamma_0}=1$ resulting in the divergence of occupation density of bosons,  see for example figure \ref{Fig4}.
The average number of bosons diverge at transition point as
\begin{eqnarray}
	2\langle a^{\dagger} a \rangle +1  \sim |\gamma_0 - \gamma|^{-\alpha}
\end{eqnarray}
with $\alpha = 0, ~ 1 $ for $\kappa=0$ and $\kappa\ne0$ respectively.

\subsection{Effective Temperature}
The response and correlation functions allows us to define a fluctuation-dissipation relationship by introducing 
distribution function $F(\omega)$:
\begin{eqnarray}
	G^K(\omega) = G^R(\omega) F(\omega) - F(\omega) G^A(\omega).
	\label{FDT}
\end{eqnarray}
This distribution function has the form  $F_{eq}(\omega)= 2n(\omega)+1={\rm coth} (\frac{\omega}{2T})$ 
with $n(\omega) = \frac{1}{e^{\beta \omega}-1}$ in thermal equillibrium. In the non-equillibrium setting here, the notion of effective temperature is determined through the low frequency 
analysis of eigenvalues of the distribution function $F(\omega)$. For our problem,
we write
\begin{eqnarray}
	F(\omega)= \sigma^z - \frac{1}{2\omega}\sum_k \frac{|g_k|^2 \omega_k}{\omega^2-\omega_k^2} \sigma^x 
	\label{temp1}
\end{eqnarray}
where $\sigma^z$ and $\sigma^x$ are Pauli spin matrices. The eigen values of e $F(\omega)$ are given by
\begin{eqnarray}
	\lambda_{\pm}(\omega) = \pm \sqrt{1+ |\frac{\Sigma(\omega)}{\omega} |^2}
	\label{temp}
\end{eqnarray}

Therefore, in the long wavelength we write  $F(\omega) \sim \frac{2T}{\omega} $.
We see from equation \ref{temp}, in this limit, eigen values $\lambda_{\pm}$ diverge as  
$\frac{1}{\omega}$. The effective temperature  $T_{\rm eff}$ is give by the dimensional coeeficient of  $\frac{1}{\omega}$ in the long wave length limit.
Therefore,  we see that $T_{\rm eff} = \gamma $ and is independent of the 
decay rate $\kappa$, cutoff frequency $\Omega$ of the bath. It can be shown true for all cases of spectral densities wit Drude-Lorentz 
cutoff. Moreover, if we chose exponential cutoff for the bath spectral density, we can show that effective temperature 
depends on cutoff frequency as well besides coupling $\gamma$.  This effective tempearture in comparison to equillibrium, 
is not an external parameter but an intrinsinc quantity
that arises due to interplay of unitary and dissipative dynamics.

\section{Fluctuations over Mean Field}
Having found out the mean field solution, we now consider the stability of these solutions 
to small fluctuations around mean field. 
We therefore add small fluctuations at tree level by 
taking $\phi_{cl} \rightarrow \phi_0 + \delta \phi$ and $\phi_q \rightarrow \delta \phi_{q}$. Therefore, from equation \ref{Sef} and
taking ${\rm O}(1/N)$ terms into account, 
we write
\begin{eqnarray}
	\tilde{S} &=& \int_{\omega} \delta \eta^{\dagger}_4(\omega) 
	\begin{pmatrix}
		0 & [\tilde{G}_{2 \times 2}^A]^{-1}(\omega) \\
		[\tilde{G}_{2 \times 2}^R]^{-1}(\omega) & \tilde{D}^K
	\end{pmatrix}
	\delta \eta_4(\omega) \nonumber \\
	&&~~~-\frac{\lambda}{2N}\int_t  \Bigg[ ( 2\phi_0 |\phi_{cl}|^2 + \phi_0^{*} \phi_{cl}^2)\phi^{*}_q  
	+ (|\phi_{cl}|^2 + |\phi_q|^2)\phi_{cl} \phi_q^{*} +{\rm c. c.}\Bigg]
\end{eqnarray}
with $\delta \eta_4(\omega) =\begin{pmatrix}
	\delta \Phi_{cl}(\omega) \\ \delta \Phi_{q}(\omega)
\end{pmatrix}
$ and 
\begin{eqnarray}
	[\tilde{G}_{2 \times 2}^R]^{-1}(\omega)
	=\begin{pmatrix}
		\omega-\omega_0 + i\kappa + \Sigma(\omega) -\frac{\lambda}{N}|\phi_0|^2 & \Sigma(\omega) -\frac{\lambda}{2N}\phi_0^2\\
		\Sigma (\omega) -\frac{\lambda}{2N} \phi_0^{*2}& -\omega-\omega_0 - i\kappa + \Sigma(\omega)-\frac{\lambda}{N}|\phi_0|^2
	\end{pmatrix},
\end{eqnarray}
while contribution to action at    ${\rm O}(\frac{1}{N})$ are due to cubic and quartic terms. 
Thus we observe that the fluctuations vanish in the thermodynamic limit $N\rightarrow \infty$.
The poles of the retarded Green’s function, give the spectrum of excitations,
while the signs of their imaginary parts determine
whether the proposed mean-field steady state is stable. A 
positive imaginary part of the spectrum implies the instability to mean field solution.
Thus, to  find the dissipative spectrum of fluctuations, we
solve ${\rm det} [\tilde{G}_{2 \times 2}^R](\omega) =0  $ and  get
\begin{eqnarray}
	\omega = - i\kappa \pm \sqrt{ (\omega_0^2 - 2\omega_0 \Sigma) -\frac{\lambda}{2N} 
		[(\phi_0- \phi_0^{*} )^2 \Sigma + 2\omega |\phi_0|^2]      } \nonumber \\
	\label{diss}
\end{eqnarray}
Therefore, in the limit of $N\rightarrow\infty$, the fluctuations are washed away, and we retain the same mean field spectrum given in equation \ref{char}. Next, we analyze the effect of fluctuations on the distribution matrix $F(\omega)$ that provides the information regarding
effective temperature. From fluctuation-dissipation relation \ref{FDT}, we can write 
\begin{eqnarray}
	F(\omega) = \sigma^z + \frac{1}{\omega} [\Sigma(\omega) - \frac{\lambda}{4N} (\phi_0^2 + \phi_0^{* 2})] \sigma^x,
\end{eqnarray}
which has the same form in thermodynamic limit $N\rightarrow \infty$ as defined in equation
\ref{temp1}. Thus fluctuations due to finite number of particles $N$ reduce the effective temperature.

Now, we take into account the contribution of cubic and quartic terms in the effective action. 
In principle we can sum up to all orders of perturbation and get the following equation
\begin{eqnarray}
	[{\rm G}_0^{-1} - \Sigma] \circ \mathcal{G}=I_{2\times 2}
\end{eqnarray}
where ${\rm G}_0^{-1}$ is the bare Greens function, $\mathcal{G}$ is the dressed Greens function due to the 
interactions and the self energy matrix is 
$\Sigma= \begin{pmatrix}
	0 & \Sigma^A \\
	\Sigma^R & \Sigma^K
\end{pmatrix} $.
However, we restrict here to the qualitative ideas, where as the full details of effects of  interactions are treated 
separately \cite{56} 
within the renormalization group approach in Keldysh space. 

We consider the effect of  fluctuations at  first order of $\frac{\lambda}{N}$. The cubic terms at this order are 
$\int_{t}[2\phi_0 \phi_{cl}^2 \phi_{q}^* + \phi_0^* \phi_{cl}^2 \phi_q + {\rm c.c.}  ]$. This term breaks the $\mathcal{Z}_2$-
symmetry, $\phi_{cl/q} \rightarrow -\phi_{cl/q}$ and can be treated as the external ``magnetic`` field  term.
In general, the fluctuations can modify the position of the critical point and these terms serve the corrections to the mean
field position of the phase transition. However, we can eliminate these odd order terms by applying the external drive.
This kind of situation also arises in the liquid-gas transition, where there is no obvious symmetry, 
however, one can choose parameters such
as density to eliminate odd terms. This phase transition,
despite the absence of symmetry, is of the Ising type \cite{55}. A similar conclusion holds 
if we take fluctuations  at higher order
of $\lambda/N$.
Moreover, we can show \cite{56} that this model undergoes a second order thermodynamic phase transition of $\phi^4$-theory with $\mathcal{Z}_2$-symmetry
We thus conclude that the driven-dissipative model considered here  undergoes a continuous Ising-
type phase transition.

\section{Conclusions}
In conclusion, we have analyzed the non-equilibrium dynamics in a long range interaction Heisenberg model coupled
to bath and driven by the dissipation at each site due to flipping of spin (spontaneous emission ). We have shown that Holstein-Primakoff transformation cannot be  faithfully applied to the entire range of the parameters of the model. In a limited domain of parameter values, we have mapped IRHM to a multimode Dicke model with non-linearities.
Using the Keldysh field theory, we have shown in the thermodynamic limit that the system boson density has a power law behavior 
with the critical exponent
depending on the values of decay constant $\kappa$ and the type of
spectral density used. 

Also that , an  effective temperature arise due to dissipation, and is shown to be depend linearly on
the effective coupling $\gamma$, 
independent of the cutoff frequency of the bath in wide class of bath spectral densities.
It is shown that  the fluctuations  due to  cubic field terms in the 
perturbation expansion violate  $\mathcal{Z}_2$-symmetry and modify the mean field critical point.
Near the steady state, however it can be shown that the dynamics is generically 
described by a thermodynamic universality class \cite{17,56} of $\phi^4$-theory of  Landau and 
Ginzburg . The emergent thermal character of driven-dissipative systems
may be expected as the quantum coherence is lost to dissipation.
\subsection*{References}

{\small \topsep 0.6ex

}

\end{document}